\documentclass[11pt,english]{article}
\usepackage{moriond_mod,epsfig}
\usepackage{amsmath} 
\usepackage{hhline}
\usepackage{amssymb}
\usepackage{times}
\usepackage{babel}
\usepackage{umlaut}
\def\Journal#1#2#3#4{{#1} {\bf #2}, #3 (#4)}

\def\PLB{{\em Phys. Lett.}  B}
\def\PRL{\em Phys. Rev. Lett.}
\def\PRD{{\em Phys. Rev.} D}
\def\ZPC{{\em Z. Phys.} C}
\def\EPJC{{\em Eur. Phys. J.} C}
\def\SJNP{\em Sov. J. Nucl. Phys.}

\def\be{\begin{equation}}
\def\ee{\end{equation}}
\def\bea{\begin{eqnarray}}
\def\eea{\end{eqnarray}}

\newcommand{\Pom}{{\ensuremath{\mathbb{P}}}}

\newcommand{\xpom}{x_{\Pom}}
\newcommand{\ftwodarg}{F_2^{D(3)} (\beta, Q^2, \xpom)}

\begin{document}
\vspace*{4cm}
\title{QCD AND DIFFRACTION AT HERA}

\author{ M.KAPISHIN }

\address{Joint Institute for Nuclear Research, Joliot Curie 6,\\
R-141980 Dubna, Russia}

\maketitle\abstracts{
Recent measurement of inclusive processes and 
hadronic final states in diffractive deep-inelastic
scattering at HERA are used to investigate the QCD factorisation
properties and study partonic structure of colour
singlet exchange. Resolved Pomeron and colour dipole models are tested by comparison 
with the data.}

\section{Introduction}

The observation of events with a large rapidity gap in the hadronic final state
at HERA~\cite{LRG_HERA}, which are
attributed to diffractive dissociation of virtual photons, led to a renewed 
interest in the study of the underlying dynamics of diffraction. Diffractive 
scattering is governed by the exchange of the Pomeron ($\Pom$), an object 
carrying vacuum quantum numbers. Pomeron exchange was introduced in Regge 
theory~\cite{Regge} to describe the high energy
behaviour of total hadron cross sections~\cite{Donnachie}, which are related to
elastic scattering through the optical theorem.

In perturbative QCD (pDCD) the vacuum exchange is modeled as a colour singlet  exchange of
at least two gluons~\cite{QCDPom} that develops into a gluon ladder between the photon and
the proton. pQCD calculations are possible
for diffractive processes where a hard scale is present: production of high
mass quarks, high momentum jets, processes with high virtuality $Q^2$ or squared 
four-momentum transfer $t$.

At HERA, by changing $Q^2$ and $t$, it is possible to vary the resolution with which the Pomeron
structure is probed in diffractive interactions, and to study its partonic content. 
HERA is thus an unique facility to study the transition from soft to hard interactions.

The kinematics of diffractive deep-inelastic scattering $ep \rightarrow eXp$
can be expressed in terms of
the variables $Q^2$,~$t$, Bjorken $x$ and the invariant mass of the $X$
system - $M_X$. In addition, the variables $\xpom$ and $\beta$ are introduced.
In the proton infinite momentum frame, $\xpom$ is the fraction of the beam
proton momentum carried by the Pomeron, $\beta$ is the fraction of the Pomeron momentum carried
by struck parton.

\section{Factorisation and diffractive parton distributions}

It has been proven that in diffractive $ep$ processes diffractive structure function
$F_2^D$ factorises into
long and short distance contributions in analogy with the inclusive $F_2$ (QCD factorisation), i.e.

\begin{equation}
F_2^{D(4)}(\beta,Q^2,\xpom,t) \sim f^D(\beta,Q^2,\xpom,t) \otimes \hat{\sigma}(\beta,Q^2)
\label{eq:QCDfact}
\end{equation}

where $\hat{\sigma}$ are cross sections for pQCD hard scattering and $f^D$ are diffractive
parton density functions (PDFs), which express conditional proton parton probability distributions
at fixed $\xpom$ and $t$. The PDFs obey the DGLAP evolution equations and are universal
for diffractive DIS processes (inclusive, jet, charm production)~\cite{QCDFact}.

If for all relevant $f^D$, the $\xpom$ and $t$ dependences decouple from the $\beta$ and $Q^2$
dependences, the extended factorisation property is known as Regge factorisation or
the "resolved Pomeron" model \cite{Schlein}:
%(see Fig.~\ref{ResPom}):   

\begin{equation}
F_2^{D(4)}(\beta,Q^2,\xpom,t) = \Phi(\xpom,t) \cdot F_2^{\Pom}(\beta,Q^2)
\label{eq:Reggefact}
\end{equation}

In this case $\Phi(\xpom,t)$ can be interpreted as the flux of the Pomeron, and $F_2^{\Pom}$ as
the structure function of the Pomeron. The flux factor describes the long distance physics at the
proton vertex, while the structure function depends on the exchanged parton densities and the
short distance physics at the photon vertex.

In the complementary "colour dipole" approach the virtual photon is considered to
fluctuate into $q\overline{q}$ or $q\overline{q}g$
states described by dipole wave functions~\cite{Dipole}. In the proton rest frame and at low
Bjorken $x$ the fluctuation happens long before the interaction with the proton. The diffractive
$\gamma^*p$ cross section is
factorised into the square of the effective dipole wave function and the square of
the cross section for the
diffractive scattering of the dipole off the proton. The dipole cross section can be calculated in
pQCD for relatively small dipole sizes, corresponding to high values of $Q^2$. There are
several "colour dipole" models which differ in
the way they treat the actual dipole-proton interaction~~\cite{Dipole,BJW,Sat,RIDI}.

\begin{figure}[!hb]
\vspace*{-0.6cm}     
\begin{center}
\begin{minipage}{0.45\textwidth}
\epsfig{file=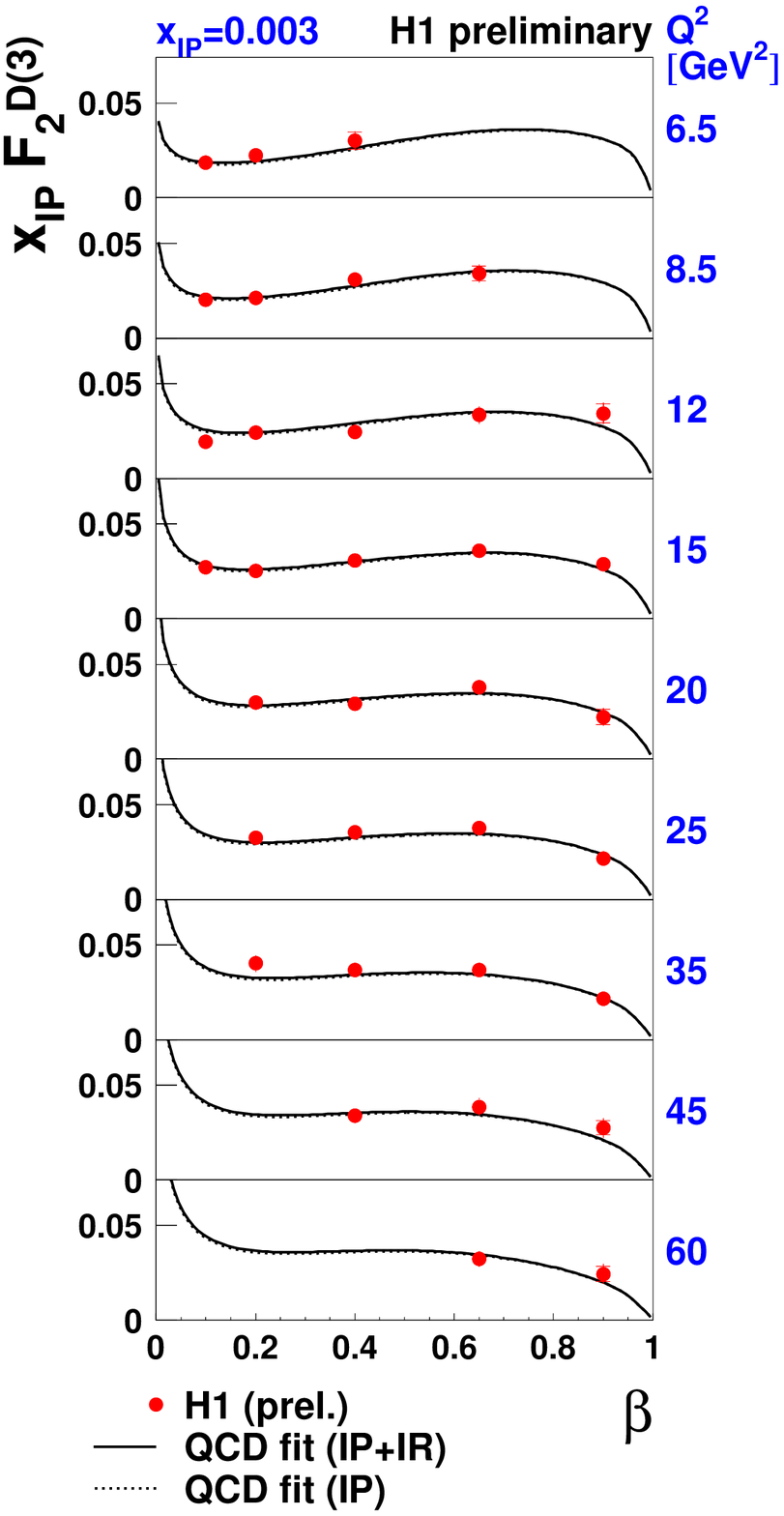,height=0.42\textheight,width=0.8\textwidth}
%\vspace{0.9cm}
\end{minipage}  
\begin{minipage}{0.45\textwidth}
\epsfig{file=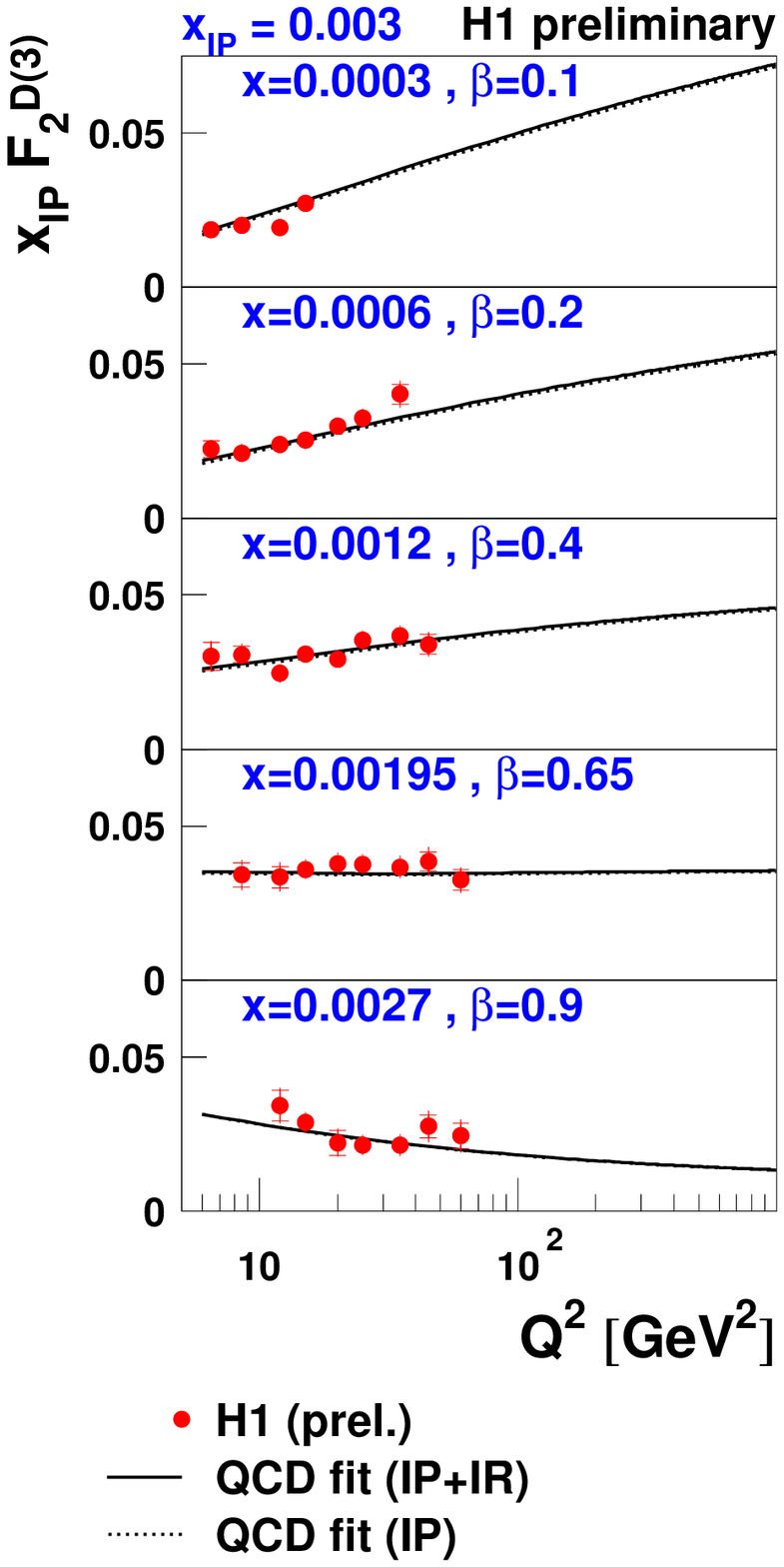,height=0.42\textheight,width=0.8\textwidth}
\end{minipage}
\end{center}
\vspace*{-0.2cm}
\caption{The diffractive structure function $\xpom\ftwodarg$ plotted as a
function of $\beta$ and  as a function of $Q^2$ at fixed value of $\xpom=0.003$. Also shown
is the result of a QCD fit to the data.}
\label{F2D3_QCD}
\end{figure}

The $\xpom, \beta$ and $Q^2$ dependence of $F_2^{D(3)}$ is studied 
in a new high precision measurement by the H1 experiment~\cite{F2D3_H1}. 
The measured structure function integrated over $|t|<1~GeV^2 \xpom
F_2^{D(3)}(\beta,Q^2,\xpom)$, 
shown in Fig.~\ref{F2D3_QCD} for a fixed $\xpom$ value, indicates a rising 
scaling violation with $\ln Q^2$, persisting up to relatively large value of 
$\beta \sim 0.65$. At the highest $\beta$, the $F_2^D$ scaling violation 
becomes negative. In this region, higher twist contributions
such as vector meson production could play a major role in the 
diffractive cross section~\cite{Sat}. 

\begin{figure}[!t]
%\vspace*{-1cm}
\epsfig{file=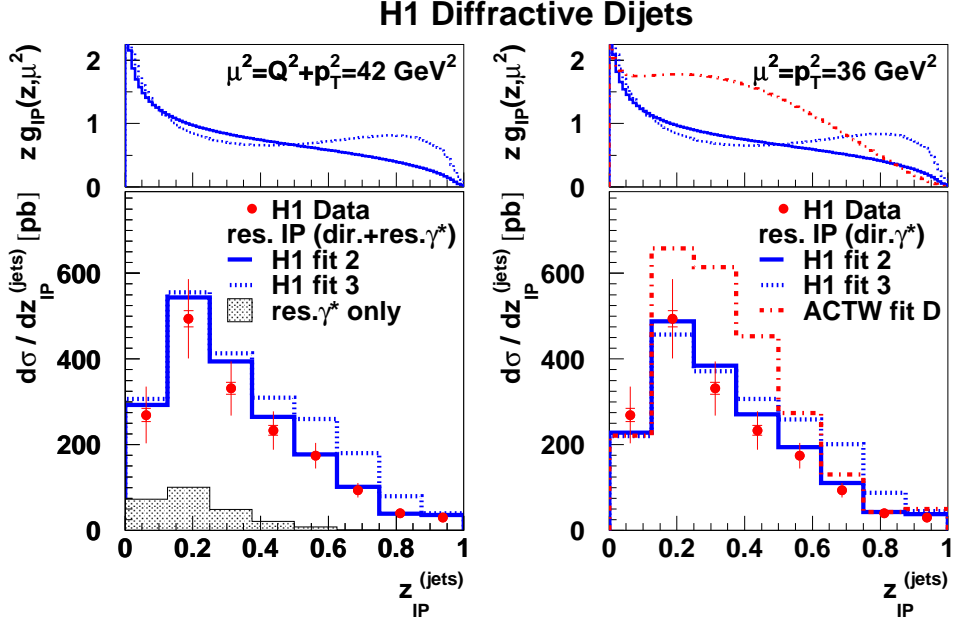,height=0.35\textheight}
\vspace{-0.2cm}
\caption{The diffractive dijet cross section as a function of $z_{\Pom}$ compared with
predictions of the "resolved Pomeron" model based on the QCD fits. The diffractive gluon
densities for each fit are shown above the data at the mean scale $p_T^2$ or $Q^2+p_T^2$.}
\label{dijets_H1}
\end{figure}

A fit assuming both QCD~\cite{QCDFact} and Regge factorisation~\cite{Schlein}
was performed in which the parton densities in
the Pomeron were evolved according to the leading order DGLAP equations. The fit result is
compared with the data in Fig.~\ref{F2D3_QCD}. The logarithmic scaling
violation in $Q^2$ and relatively flat $\beta$ dependence is described by the fit, which predicts a
partonic momentum distribution of the Pomeron dominated by a gluon contribution extending
to large fractional momenta.

\section{Diffractive dijet and charm production}

According to the QCD factorisation theorem~\cite{QCDFact}, parton distributions extracted from
the QCD fits to inclusive diffraction can be used to describe to hadronic final states in
diffractive DIS
for the same $\xpom$ and $t$. Of particular interest are
measurements of diffractive dijet and open charm production, since the implied boson
gluon fusion process $\gamma^*g \rightarrow q\overline{q}$ provides a direct probe of the gluon 
content of the Pomeron. The presence of the hard scale, provided by the high momentum jets or 
mass of the charm quark, allows a variety of perturbative QCD-based models of diffraction
to be tested.

In Fig.~\ref{dijets_H1} the quantity $z_{\Pom}^{jets}$ representing the
fraction of the hadronic energy in the final state contained in the two jets of a high
$p_T$ jet sample, is compared with the
prediction from the "resolved Pomeron" model~\cite{Schlein} based on different sets of Pomeron
gluon distributions obtained from the leading order DGLAP fit of earlier $F_2^{D(3)}$
data by the H1 experiment~\cite{H1_LRG} and others~\cite{ACTW}.

The data favour the Pomeron dominated by gluons with a gluon momentum distribution that is
relatively flat in $z_{\Pom}^{jets}$. For diffractive scattered  $q\overline{q}$ photon
fluctuations, a distribution peaked at $z_{\Pom} \sim 1$ is expected. The low values of $z_{\Pom}$
correspond to dominance
of $q\overline{q}g$ over $q\overline{q}$ scattering. The $\xpom$ and $\beta$ distributions of
the dijet cross section are in agreement with the "resolved Pomeron" model with the Pomeron
intercept $\alpha_{\Pom}(0)$ extracted from inclusive $F_2^{D(3)}$ data~\cite{H1_LRG}. In the
"colour dipole" approach the model which allows non-ordered $k_T$ distribution of
the partons~\cite{BJW} 
gives good description of the dijet data whereas the model with strong $k_T$ ordering 
($k_T^g~\ll~k_T^q)$~\cite{Sat} is much below the data. The result indicates that non-ordered $k_T$
contributions could be important in diffractive dijet production.

The "resolved Pomeron" model with various assumptions for the partonic composition of the
colourless exchange~\cite{H1_LRG,ACTW} provides a reasonable description of the diffractive
$D^*$ production measured by the H1 and ZEUS experiments~\cite{H1_Dstar,ZEUS_Dstar}. Predictions of 
two gluon exchange "colour dipole"  
models~\cite{BJW,Sat,RIDI} match the H1 data at low $\xpom$ and describe the ZEUS data except for 
 the $\beta$ distribution. Higher statistics of diffractive $D^*$ events are needed to
distinguish
between the models.

\section{Event shapes and three jet production}

\begin{figure}[!htb]
\vspace{-1.5cm}
\begin{center}
\begin{minipage}{0.45\textwidth}
\epsfig{file=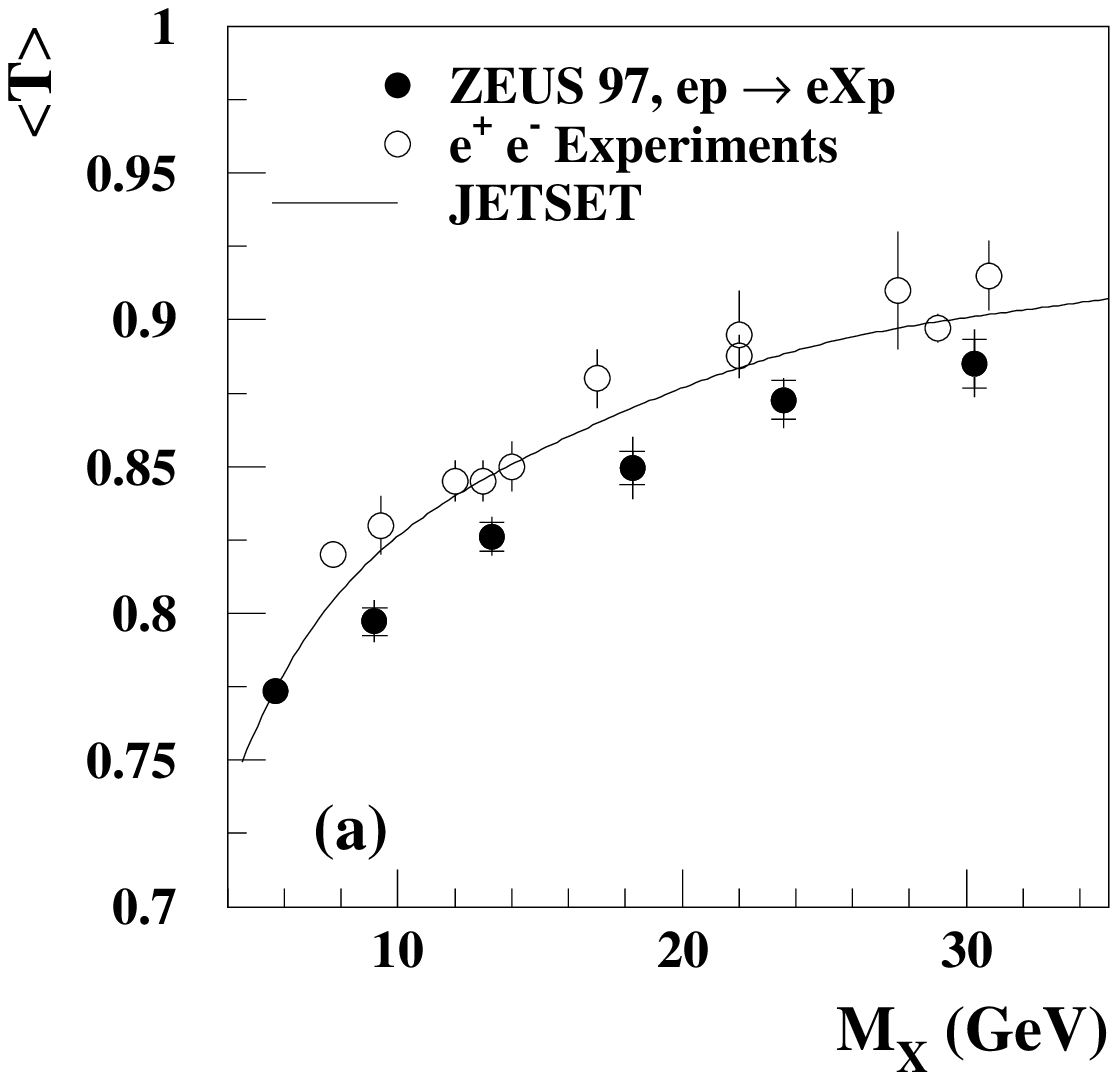,height=0.4\textheight,
       bbllx=210pt,bblly=325pt,bburx=570pt,bbury=745pt,clip=}
\end{minipage}
\begin{minipage}{0.5\textwidth}
\epsfig{file=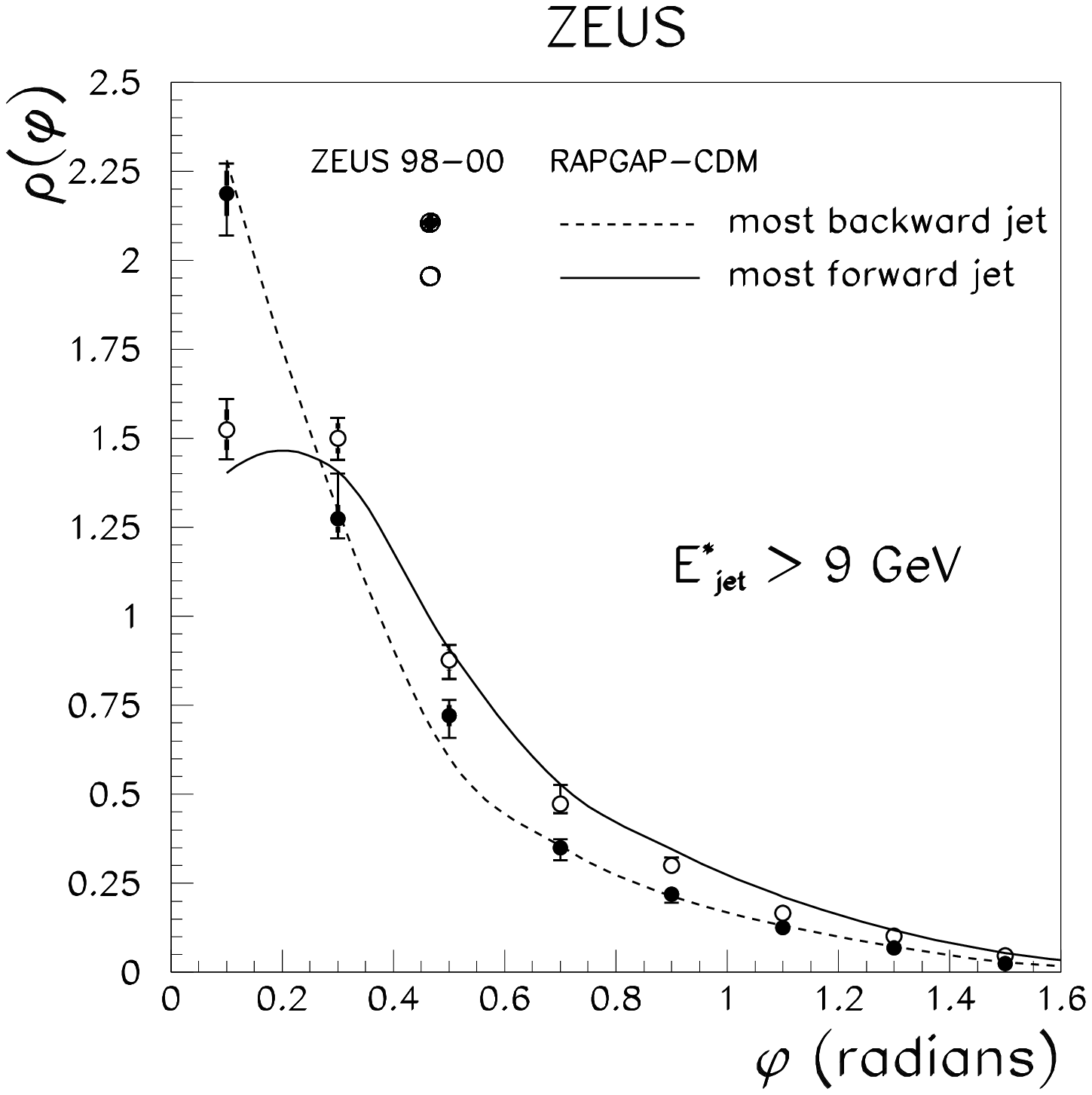,height=0.4\textheight}
\end{minipage}
\end{center}
\vspace{-0.8cm}
\caption{{\bf Left:} Average thrust $\langle T \rangle$ of the diffractive DIS hadronic final state
as a function of $M_X$. {\bf Right:} The differential jet shape $\rho(\phi)$ for the most-forward
and most-backward jets in three-jet events, where the Pomeron defines the forward direction}
\label{ZEUS_HFS}
\end{figure}

The properties of the diffractive hadronic final state were studied recently by the ZEUS
experiment in terms of 
global event-shape variables such as thrust in the center-of-mass (CMS)
frame~\cite{ZEUS_HFS}.
For collimated two-jet $q\overline{q}$ events the value of thrust approaches 1, while events with
an isotropic shape yield values close to 0.5. $q\overline{q}g$ final states are therefore
expected to yield thrust values lower than $q\overline{q}$. In
Fig.~\ref{ZEUS_HFS}(left) the thrust
distribution is shown as a function of the invariant mass $M_X$ of the hadron final state.
The diffractive hadronic final state becomes more collimated as $M_X$ increases, a tendency also 
observed in $e^+e^-$ annihilation. The diffractive
events show a thrust distribution which is shifted to low values compared with processes 
$e^+e^-\rightarrow q\overline{q}\rightarrow hadrons$, indicating that they are more
isotropic. This can be attributed to contributions not present in $e^+e^-$ annihilation, such as the
boson-gluon fusion process in the "resolved Pomeron" approach~\cite{Schlein}, or $q\overline{q}g$
production from the dissociation of the virtual photon in the "colour dipole"
approach~\cite{Dipole}.

Diffractive three jet events were studied by the ZEUS
experiment~\cite{ZEUS_3jets}. The
differential jet shape $\rho(\phi)$, defined as the fraction of the jet energy which lies
inside annulus at angular distance $\phi$ around the jet axis, is shown in 
Fig.~\ref{ZEUS_HFS}(right) as a function of $\phi$ for the
most-forward and most-backward jet. The forward direction is defined by the Pomeron, the
backward direction - by the virtual photon. The jet in the Pomeron direction is broader that the
jet in the photon direction. This measurement supports the picture where three jet final
state is dominated by $q\overline{q}g$ configuration with a gluon emitted in the Pomeron direction. 
The jet shape distribution is reproduced by the "resolved Pomeron" model with the Pomeron dominated
by gluons~\cite{Schlein,H1_LRG}.   

\section{Summary}

A consistent picture of inclusive diffraction and hadronic final states is observed in DIS at HERA.
The $F_2^D$ and dijet measurements support QCD factorisation with diffractive parton density
functions dominated
by gluon contribution. Event shape distributions and the topology of three jet events also
indicate a significant
contribution of $q\overline{q}g$ systems in final state the with gluon emitted in the Pomeron
direction. A variety of diffractive DIS data are successfully described by the "resolved
Pomeron" model
in the proton infinite momentum frame and "colour dipole" models in the proton rest frame.

\end{document}